\documentclass[conference]{IEEEtran}

\ifCLASSINFOpdf
\else
\fi
\usepackage{graphicx}
\usepackage{grffile}
\usepackage{epstopdf}
\usepackage{amsmath}
\usepackage{amssymb}
\usepackage{amsthm}
\usepackage{subfig}
\usepackage{color}
\usepackage{multirow}
\usepackage[font=footnotesize,labelfont=]{caption}

\begin{document}
%
\title{Making a Dynamic Interaction Between Two Power System Analysis Software}

\author{\IEEEauthorblockN{Amirreza Sahami}
\IEEEauthorblockA{Department of Electrical and\\Computer Engineering\\
UNC Charlotte\\Charlotte, USA\\
asahami@uncc.edu}
\and
\IEEEauthorblockN{Shahram Montaser Kouhsari}
\IEEEauthorblockA{Department of Electrical and\\Computer Engineering\\
Amirkabir University of Technology\\
Tehran, Iran\\
smontom@aut.ac.ir }}


\maketitle

\begin{abstract}
Different goals in studying a piece of equipment or a phenomenon, and the need for different accuracies in studies and manufacturing new instruments, persuade engineers to use different mathematical models for studying the behavior of phenomena and elements. On the other hand, calculation complexity in the study of large systems, and the rising costs of industry in conjunction with development of capabilities of new generations of computers, have made computer simulation an unavoidable choice. There are different software applications in various fields of electrical engineering, and each one has its own specific features and usages. If we could use features of different software simultaneously, then we could improve the accuracy of analyzing engineering complexities. In this paper, we have used mapping to create dynamic interaction between two existing software packages, PASHA and MATLAB, which are popular software among electrical engineers. By linking these two software packages, we may use MATLAB and Simulink capabilities in PASHA software.
\end{abstract}


%
\IEEEpeerreviewmaketitle

\section{Introduction}
Prior to the era of digital computers, analogue computers were used to simulate dynamic engineering problems, which were modeled using differential equations. In late 1950s, after the appearance of new digital computers, the first software packages were developed to calculate short circuit capacities and to perform transient stability studies. In these programs, generators were modeled by a constant voltage source in series with a variable reactance and the controllers' effects were ignored.

The improvement of digital computers in the late 1960s helped to develop better software, which brought more details of the generators' models and their controllers into consideration. 

Since then, much progress has been made in modeling power system equipment, and IEEE committees have provided various standard models for equipment, which have been used to develop different power system software. However, the invention of new controllers brought some complaints about the lack of accuracy of the provided standard models. Also, in some studies it was shown that the models used for machines were not accurate enough \cite{teze kouhsari}.

Until 1977, a large number of researchers were focused on studying the power system dynamics to gain a better perception about generators' dynamic behaviors to provide more detailed and accurate models \cite{teze kouhsari}.

The behavior of modern interconnected power systems needs to be predicted more accurately due to its complexity and the significant economic impacts and security consequences that might happen in case of a failure. Therefore, we need to model our equipment in accordance with up-to-date models, and we need tools with the ability to flexibly model new equipment. Fortunately, fast computers help engineers to analyze and predict more complex situations than was possible before \cite{kundur,teze lari}.

Different software packages are designed to fulfill different goals, and each of them has its own pros and cons. For instance, EMTP simulates the power systems in time domain, which is mostly effective in studying fast transient phenomena, and it is not recommended for performing load flow or transient stability analysis because of time concerns. On the other hand, some software packages such as Power Apparatus and System Homological Analysis (PASHA) use frequency domain equations to simulate networks, which decrease the simulating time for load flow and stability studies; however, it is not suitable for fast transient studies \cite{site pasha}. These kinds of limitations in software packages have led engineers to think about connecting software packages and using the better capabilities of each package in the other ones. Hence, there has been a growing trend in software development toward open systems, in which each part of the software is designed based on a protocol for easier interaction. Having the ability of dynamic data interaction in operating systems, such as Windows, and inter-medium processing in Solaris and Unix, has made it easier to connect different software packages \cite {open systems,9 ghadim,10 ghadim,20 ghadim}.

Widely used software packages, such as Labview, PSCAD, and EMTDC, have been designed with the ability to take advantage of the noted capabilities of operating systems. MATLAB has a similar capability, and it is already connected to EMTP and NEPLAN. ASPEN, which is used in petrochemical studies, is another software that has been connected to MATLAB \cite{20 ghadim,site neplan,site aspen}.


PASHA is connected to LabView, EMTP, and DigSilent. However, it should be noted that it cannot dynamically interact with them, which means that we cannot simulate part of the network in PASHA and other parts in DigSilent and run the simulation \cite{site pasha}.
 
PASHA is strong in simulating the real industrial power networks. However, it lacks toolboxes such as the control toolbox or the Fuzzy toolbox. On the other hand, MATLAB is developing quickly and provides new capabilities and toolboxes in its new versions, which in turn provides us the ability to model newer equipment. However, MATLAB is not an adequate software for simulating real power systems compared with professional software packages, such as PASHA or PSSE. Hence, considering all the works in this field and the growing need for strong and accurate power system simulators, it would be valuable to connect PASHA and MATLAB to each other so that they can have a dynamic data interaction in order to use the MATLAB capabilities in PASHA \cite{20 ghadim,21 ghadim,23 ghadim}.

There are several benefits in undertaking this research project. Although power systems can be modeled and analyzed in MATLAB, there is no comprehensive power system analysis toolbox in MATLAB for large networks. Also, professional power system analysis software packages have been specifically designed to simulate power system networks. These professional software packages usually have a lower simulation time, and are easier to use while drawing and analyzing industrial power networks in comparison with MATLAB. In contrast, MATLAB has a large library of different control functions, and due to its variety of toolboxes, it has numerous predefined, easy-to-use algorithms. Therefore, MATLAB can be a good option while trying to find a new control strategy for applying to an industrial network. The strong mathematical capabilities of MATLAB, combined with the fast and highly technical industrial modeling capabilities of PASHA, represents a comprehensive tool for electrical engineers. However, the simulation speed decreases in combined simulation (i.e. having part of the network in one software and the rest in the other one). Therefore, it is preferred to use combined simulation as a medium step to find the right algorithms and controllers between different options. Afterwards, to achieve the highest simulation speed, the entire module would be modeled in a professional software.

This article is organized as follows: First, UDEM is introduced. It follows with the proposed method for connecting MATLAB and PASHA dynamically. Finally, the results are represented for validation tests.


\section{UDEM AND PASHA}
 User Defined Equipment Modeling (UDEM) is part of the PASHA software package. UDEM can be used as an independent simulation package; in this respect, it is analogous to analogue computers \cite{teze kouhsari}. It can also be attached to another power system simulator like PASHA. The most important role of UDEM is to provide the capabilities of graphically drawing simple transfer functions, radial networks, and to model basic equipment. It can be likened to the "Commonly Used Blocks" library of Simulink \cite{teze lari}. 

\subsection{UDEM Working logic}
Before a module in a software like UDEM can be used by simulating routines, it needs to be interpreted as a set of mathematical equations that define a relation between inputs and outputs of the module. There are three ways to do so \cite{teze kouhsari,teze lari}:

\begin{enumerate}
\item 
Considering the entire block-diagram of the model and finding one transfer function that can present the effect of all the blocks.
\item 
Finding the state space equations for each block and solving them simultaneously.
\item 
Considering each block as a separate unit and finding its outputs based on its inputs.
\end{enumerate}

The first solution seems the most attractive one. However, developing a code that can always find the right answer is difficult, especially while having nonlinear elements in models.

 \begin{figure}[t!]
	\centering
	\includegraphics[clip,width=\columnwidth]{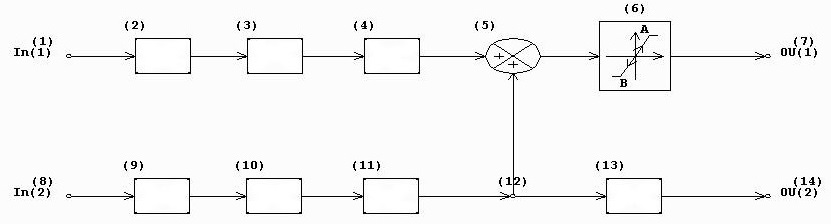}%
	\caption{A sample block diagram in UDEM}
	\label{Figure1}
\end{figure}

Although the second and third solutions are similar, using the second method requires matrix algebra, which makes the solution complex for non-linear elements' models. Consequently, the third method seems more practical. Based on the third solution, each block is considered as an independent unit and calculates the output based on its input as shown in Fig.\ref{Figure1}. In this module, two inputs and outputs are considered. The output of each element can be gained based on its input. Hence, if the input values are (IN(1) and IN(2)), output values can be calculated. First, the outputs of elements that are directly connected to the inputs (Elements number 2 and 9) are calculated. Then, the outputs of these two elements are considered as inputs for the next elements and so forth. Therefore, the required data for simulation routines are:
\begin{enumerate}
	\item 
	The relation between input and output of each block.
	\item 
	The connection between the blocks.
\end{enumerate}

Hence, for simulating a model, instead of considering one large model, several small blocks with their own operation can be considered. Then the output of each element during transient or steady-state studies would be found.

\section{A REVIEW ON MEMORY MAPPING IN MATLAB}
Memory mapping is a process that maps all or parts of a file on a disk in a range of addresses in a software's address space. The software can access the file just like it accesses the dynamic memory, which makes it faster than using commands, such as ``fread" and ``fwrite", for reading and writing the data \cite{site matlab,22 ghadim}.
Using memory mapping gives another advantage for providing access to a file's data by using some standard Indexing Operations. When a file is mapped, for reading and writing its data, the same commands that are used in the MATLAB workspace can be applied. The contents of the mapped file seem like an array in the current working space, which can be easily read from or written to this array \cite{site matlab}.
\subsection{Benefits of Memory Mapping}
Faster file access, efficiency, and sharing memory between different software packages and applications are the main advantages of using memory-mapping \cite{site matlab}.
\begin{itemize}
	\item Faster File Access
		
	 Memory Mapping is faster because data can be read and written using the capabilities of virtual memory, which is an internal feature of operating systems, instead of being processed in the buffer first.
	 
	 \item Efficiency
	 
	 The mapped files provide an ability that software can access data in a large file without having to read the entire file.
	 
	 \item Sharing Memory
	 
	 Mapping gives the ability to share data between different software packages and applications. It is possible to use this feature to connect MATLAB to other software packages.
\end{itemize}

The size and format of the file, the system platform, and the manner of using the data determines the impact of memory-mapping.
It is most useful for binary files in the following situations:
\begin{itemize}
	\item For using large files more than once
	\item For writing small files in the memory frequently
	\item For sharing data
	\item For the data in array format in MATLAB
\end{itemize}

Files larger than about two hundred mega bytes use a vast amount of virtual addressing space used by MATLAB, and it can cause an ``out of memory" error to be generated by MATLAB \cite{20 ghadim}.

\section{CONNECTING UDEM AND MATLAB}

\subsection{Sharing File and Memory}
File mapping can be used for sharing a file or memory between different processes in a computer. Sharing a file is not beneficial for dynamic interaction due to low speed. Instead, data is mapped into the memory. Mapping a file makes a specific part of a file visible in a section of the memory of the process that would use that data. For files that are larger than the address of the mapping, only a part of the file can be mapped. After completing this process, the mapping can be cleaned, and the rest of the file can be mapped. The mapped data is stored in the system ``paging files" \cite{site neplan,23 ghadim,site matlab,22 ghadim,site microsoft}.

To share data by this method, all the processes should use the same name, the same handle, or the same file mapping object.

\subsection{Introducing Employed Functions}

In order to share data between software packages by data mapping, WinAPI functions are employed. A brief explanation about each function is as follows \cite{site matlab,22 ghadim,site microsoft}:

\begin{itemize}
	\item CreateFileMapping Function: \\ This function creates or opens a named or unnamed mapping for a specific file. After a mapping is created for a file, the size of the file should not be greater than the mapping size. Otherwise, all the content of the file cannot be shared.
	
	\item OpenFileMapping Function: \\
	This function opens a named mapping.
	
	\item MapViewOfFile Function: \\
	This function maps a view of a file mapping into a section of the memory of the process that will use that data.
	Once a mapping is backed with a virtual paging file (i.e. using the ``CreateFileMapping" function with ``INVALID\_HANDLE\_VALUE" as its parameter), ``PagingFile" should be a sufficient size to include the entire mapping or else the ``MapViewOfFile" function will not work properly. The predefined initial conditions in a ``PagingFile" are zero.
	\item CloseHandle Function
	
	Mapping a file will cause the virtual address spaces to become unavailable for further allocation. To free that address, the ``CloseHandle" function should be used after deleting that mapping via the ``UnmapViewOfFile" function. The ``CloseHandle" function closes an open object handle.
	
	\item UnMapViewOfFile Function:\\
	This function deletes the view of a file mapping in the memory and prepares it for other applications.
	

\end{itemize}

  \begin{figure}[t!]
	\centering
	\includegraphics[clip,height=4.35cm,width=\columnwidth]{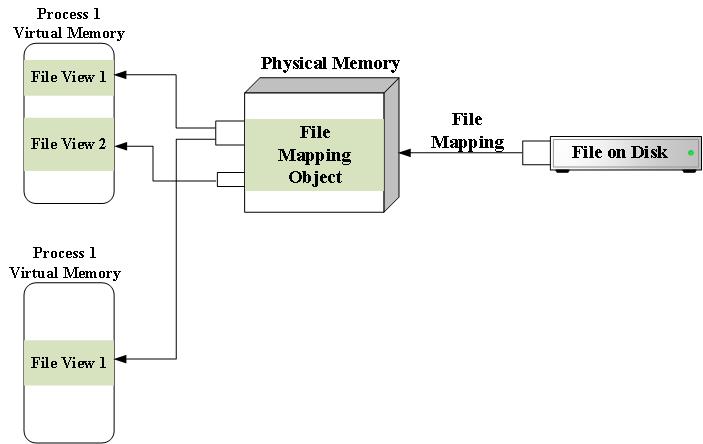}%
	\caption{Schematic about how sharing memory works}
	\label{Figure2}
\end{figure}

\subsection{Create a file Mapping Object }
The ``CreateFileMapping" function returns a handle, which is used when creating a file view, and it can access the data in the shared memory. When the ``CreateFileMapping" function is employed, the name for the mapping, the number of the bytes that will be mapped, and the reading and writing permission should be determined. The first process that uses the ``CreateFileMaping" function creates the mapping. However, when another process uses this function with the name of an existing mapping, it will receive a handle of that mapping \cite{site matlab,site microsoft}.
It should be noted that creating a mapping does not occupy any physical memory, but just allocates it. A schematic view about memory sharing is depicted in Fig.\ref{Figure2}.

The easiest way to receive the handle created by a process for other processes is by using the ``OpenFileMapping" function and defining the name of the created mapping. This type of memory sharing is called "Named Shared Memory." The other type is ``Unnamed Shared Memory." 
A process that shares a file memory needs to employ ``MapViewOfFile" function to see the file contents. It should be considered that a mapping will remain in the memory until all the applications that are using it close the handle related to that mapping \cite{site matlab,site microsoft}.

 \begin{figure}[b!]
	\centering
	\includegraphics[clip,width=\columnwidth]{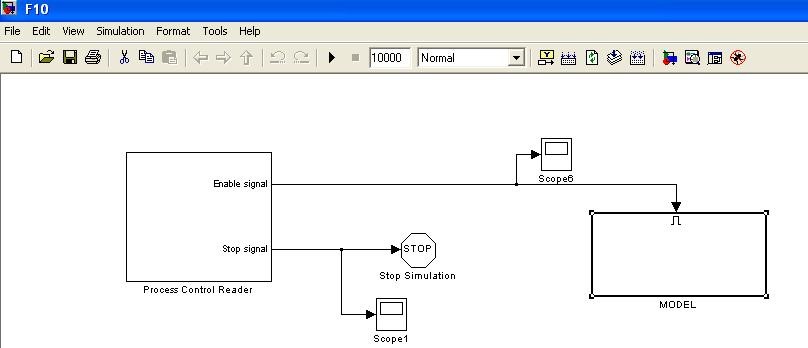}%
	\caption{The main Pre-Built Simulink File. User’s model should be copied in the designated area}
	\label{Figure3}
\end{figure}

	
	When using the ``Named Shared Memory" method, the first process creates a file mapping via the ``CreateFileMapping" function with a name and the function parameter should be defined as ``INVALID\_HANDLE\_VALUE." By defining the second parameter as ``PAGE\_READWRITE," the process will be able to read and write to the file even when the mapping is being viewed. Then, this first process uses the file mapping object (which is created by the ``CreateFileMapping" function and is returned in response to the ``MapViewOfFile" function) in order to view the contents of a file in an address. When this process no longer needs the mapping data, it should close the mapping via the ``CloseHandle" function. When all the handles are closed, the system can free the parts of the paging files that were used.
	
	
	The second process can access the data written by the first process via the "OpenFileMapping" and the "MapViewOfFile" functions \cite{site microsoft}.

\section{LINKING UDEM AND MATLAB}

Linking PASHA and UDEM with MATLAB and Simulink has been achieved using virtual memory and applying the functions noted earlier. About 17,000 lines of codes have been developed in programming languages C, Fortran, MATLAB, and WinAPI. 

Twenty three blocks in UDEM can interact with MATLAB and Simulink without any limitation in data type. If the blocks used in Simulink are not memory blocks that need to be initialized, such as integrators, each block can be used several times. Functions defined in each block of UDEM can be used independently. This way, we can use each block in each model up to 30 times. For example, if a specific block is used in up to 20 modules, and it is used 30 times in each module, then 1800 data inputs from PASHA are sent to MATLAB and 600 outputs of MATLAB are received back in PASHA. However, because most of the typical models in Simulink include memory blocks, using the maximum capacity of the interaction block is not usually possible.

 \begin{figure}[t!]
	\centering
	\includegraphics[clip,height=4cm,width=\columnwidth]{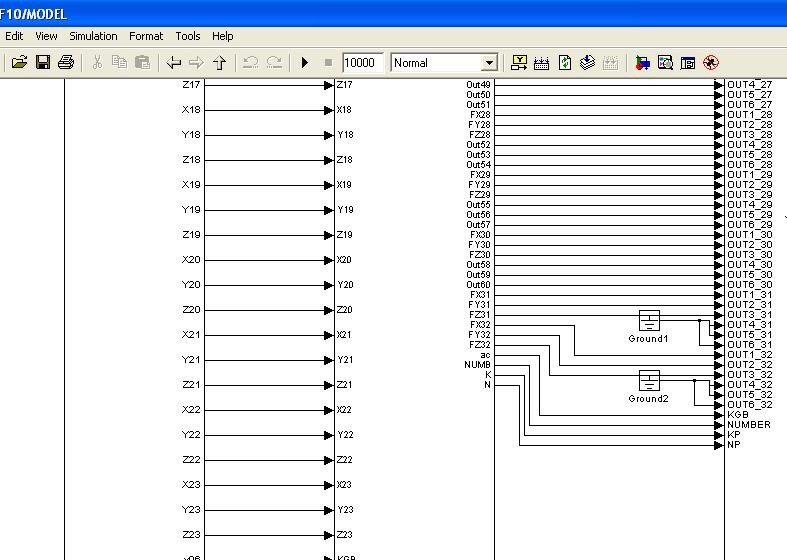}%
	\caption{A view of inside Process Control Block which controls the dynamic data interaction between PASHA and MATLAB}
	\label{Figure4}
\end{figure}

In the UDEM drawing part, the function block can be used to apply a function from UDEM's internal library \cite{teze kouhsari,teze lari} or to send the data to a Simulink model and use MATLAB and Simulink capabilities. To do so, after connecting the desired inputs and outputs to the function blocks in UDEM, the desired Simulink model should be made as such that it can dynamically interact with UDEM. In order to achieve the dynamic interaction, the desired Simulink model must be copied in a pre-created Simulink file and the suitable inputs and outputs of the model in Simulink should be connected according to the name of the functions that have been defined in UDEM.

In Fig.\ref{Figure3} the left block is a controller that checks the flags sent between PASHA and MATLAB, and the right block includes three sub-blocks that are shown in Fig. 4.
\begin{table}[b!]
	\centering
	\caption{Generator Data}
	\label{gendata}
	\begin{tabular}{|c|c|c|c|}
		\hline
		V-NOM(kV) & 13.8 & DA-TR-TC(P.U.) & 8.3200 \\ \hline
		RES(P.U.) & 0.07157 & DA-ST-X(P.U.) & 0.4455 \\ \hline
		REAC(P.U.) & 4.45513 & DA-ST-TC(P.U.) & 0.0250 \\ \hline
		ZSQ-R(P.U.) & 948.718 & Inertia Constant & 2.4922 \\ \hline
		ZSQ-X(P.U.) & 0.30449 & Gen-MW & 20 \\ \hline
		DA-TR-X(P.U.) & 0.5897 & Gen-MVAR & 20.82 \\ \hline
	\end{tabular}
\end{table}
 \begin{figure}[b!]
	\centering
	\includegraphics[clip,height=4cm,width=.4\columnwidth]{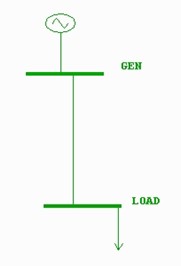}%
	\caption{A simple test network}
	\label{Figure5}
\end{figure}

 \begin{figure}[!h]
	\centering
	\includegraphics[clip,height=4.35cm,width=\columnwidth]{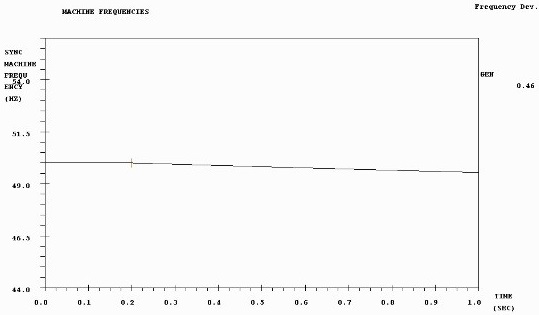}%
	\caption{Network Frequency without using interaction block}
	\label{Figure6a}
\end{figure}
\begin{figure}[!h]
	\centering
	\includegraphics[clip,height=4.35cm,width=\columnwidth]{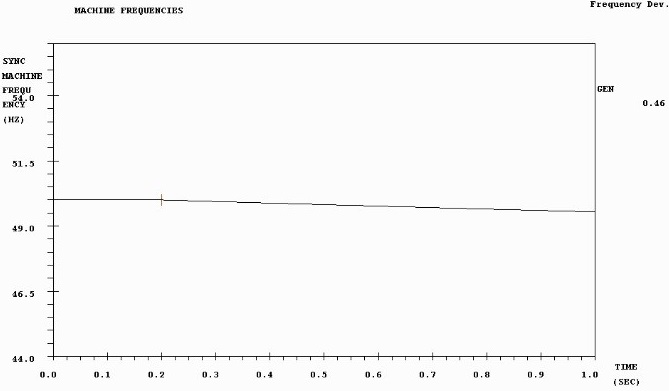}%
	\caption{Network Frequency using interaction block}
	\label{Figure6b}
\end{figure}
\begin{figure}[!h]
	\centering
	\includegraphics[clip,height=4.35cm,width=.94\columnwidth]{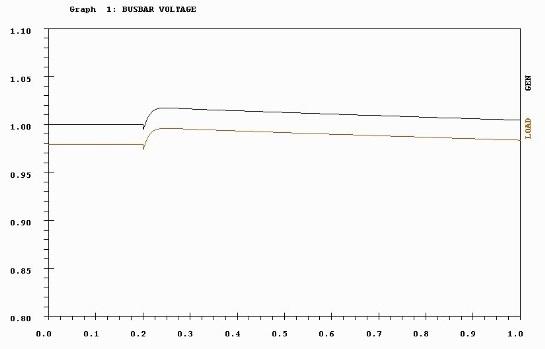}%
	\caption{Bus Voltage without using interaction block}
	\label{Figure7a}
\end{figure}
\begin{figure}[!h]
	\centering
	\includegraphics[clip,height=4.35cm,width=.94\columnwidth]{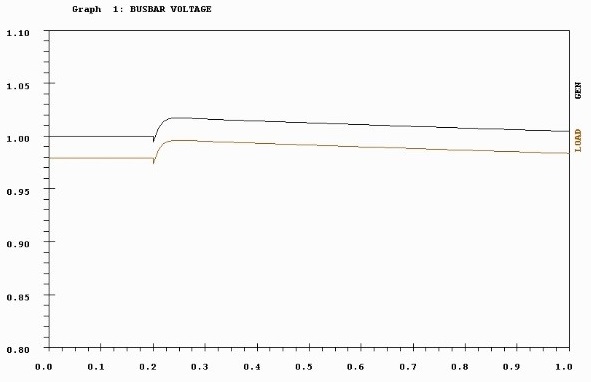}%
	\caption{Bus Voltage using interaction block}
	\label{Figure7b}
\end{figure}
\begin{figure}[h!]
	\centering
	\includegraphics[clip,height=4cm,width=.4\columnwidth]{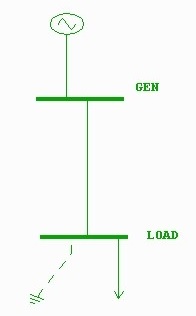}%
	\caption{Test System. A three phase fault is applied to LOAD BUS}
	\label{Figure8}
\end{figure}
\begin{figure}[h!]
	\centering
	\includegraphics[clip,height=4.35cm,width=.94\columnwidth]{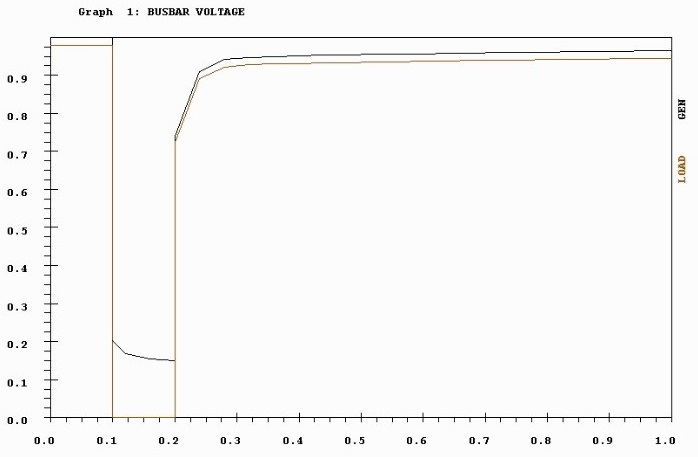}%
	\caption{Bus Voltage without using interaction block}
	\label{Figure9a}
\end{figure}
\begin{figure}[h!]
	\centering
	\includegraphics[clip,height=4.35cm,width=.94\columnwidth]{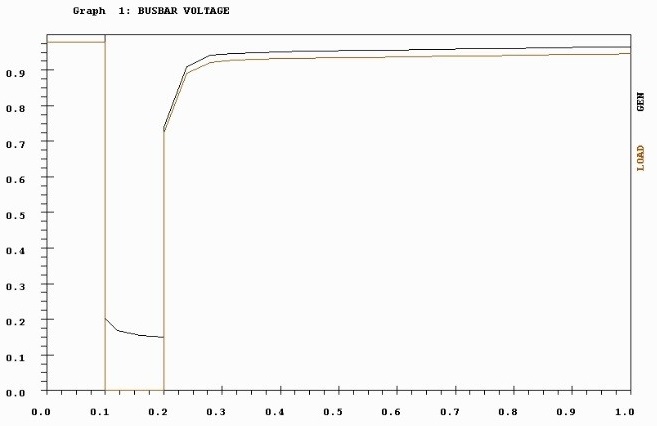}%
	\caption{Bus Voltage using interaction block}
	\label{Figure9b}
\end{figure}
\begin{figure}[h!]
	\centering
	\includegraphics[clip,height=4.35cm,width=.94\columnwidth]{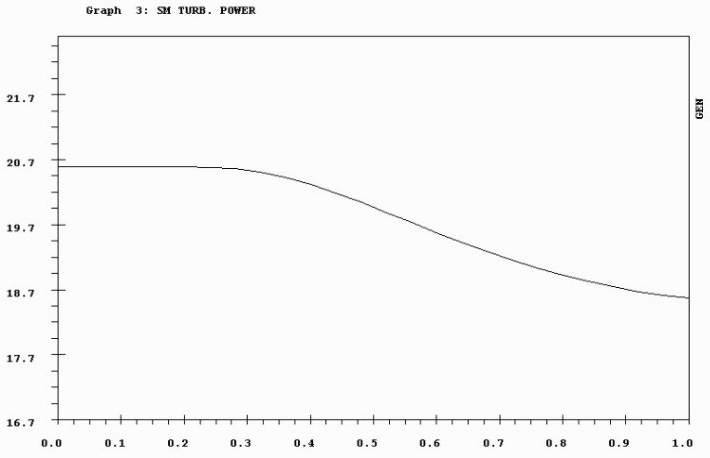}%
	\caption{Turbine Power without using interaction block}
	\label{Figure10a}
\end{figure}
\begin{figure}[h!]
	\centering
	\includegraphics[clip,height=4.35cm,width=.94\columnwidth]{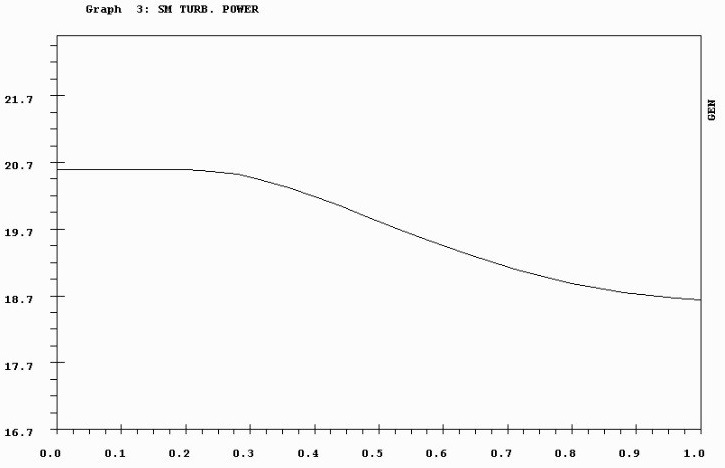}%
	\caption{Turbine Power using interaction block}
	\label{Figure10b}
\end{figure}

In Fig.\ref{Figure4} the left block reads the data and some control variables sent from PASHA. The middle block includes the Simulink models that receive the required UDEM output data from the left block and considers them as inputs for the Simulink models. The right block returns the outputs of the Simulink models to UDEM and this cycle is repeated in each time step of the simulation.

It is worth mentioning that a special m-file has been developed that the user must run after putting the Simulink models in the provided main Simulink file. This m-file does the required process for initialization of the system and creates an executable file named MAT.exe that will be used once the program is run.

\begin{table}[t!]
	\centering
	\caption{Comparing Simulation Time in the Noted Tests}
	\label{time}
	\begin{tabular}{|c|c|c|}
		\hline
		Simulation Time & \begin{tabular}[c]{@{}c@{}}Without Using the\\  Data Interaction Block\end{tabular} & \begin{tabular}[c]{@{}c@{}}With Using \\ Data Interaction Block\end{tabular} \\ \hline
		Test1           & 0.23                                                                                & 1.94                                                                         \\ \hline
		Test2           & 0.31                                                                                & 2.64                                                                         \\ \hline
	\end{tabular}
\end{table}

After running PASHA or UDEM, the initialization process starts and then routines related to the dynamic data interaction with MATLAB will be run. Then a function transfers the data from PASHA or UDEM to MATLAB. Here, PASHA can be considered the first process in the discussion made in section IV. This function pours the 
data in a pointer and, with respect to the current flag in the pointer, maps the data into the memory. Then the flag changes, and MATLAB recognizes the flag change. MATLAB can now be considered the second process in the discussion made in section IV. MATLAB runs the simulation for one time step, puts the data in a mapping, and sends a message to UDEM by changing the flag. Next, UDEM starts reading MATLAB outputs from the memory (at this phase of the interaction, MATLAB is like the first process and PASHA is like the second process according to discussion of section IV) and can use them for the next time step of simulation, which means one time step of simulation has been done in both PASHA and MATLAB. This process repeats to the end of the simulation time.
There are few limitations from MATLAB in making executable files. For example, the model should not include algebraic loops or level 2- MATLAB s-functions\cite{site matlab}. These limitations are explained further in PASHA manuals.

\section{TESTS AND RESULTS}

To test the authenticity of this Interaction-Block, more than 300 different tests have been conducted. However, only a few are presented here due to lack of space.

Fig.\ref{Figure5}. shows the picture of a simple single machine network. Machine data can be found in Table.\ref{gendata}. 
System load is 20 MW and 20 MVAR. At 0.2s, a 5MW load is switched in. The generator is equipped with an AVR and a governor. The generator responses are depicted for two states: a) where all the network and its controllers are in PASHA and b) where part of the controllers are modeled in UDEM and part of it is modeled in MATLAB, and the designed interaction-block is used to make the model complete. Result are shown in figures 6-9.

The second test is applied on the same network. At t= 0.1s, a three phase fault is applied to the LOAD bus. The fault is removed at t = 0.2s. The system is shown in Fig.\ref{Figure8}.  Results can be seen in figures 11-14.

Table.\ref{time}. shows the simulation time for the noted states.

\section{Conclusion}
In this paper, the history and importance of computer simulation in power system engineering is discussed. The necessity of connecting different software packages to use their capabilities in each other is explained, and it is shown that by using memory mapping, a dynamic interaction between software packages can be provided. The "Named Shared Memory" method is applied to two software packages, PASHA and MATLAB, and the results are shown. This method can be used without losing any data accuracy.


\end{document}